\documentclass[aps,onecolumn,groupedaddress,showpacs,floats]{revtex4}
\usepackage[dvips]{graphicx}
\usepackage{bm}
\begin{document}

\title{Detecting Super-Counter-Fluidity by Ramsey Spectroscopy}
\author{Anatoly Kuklov}
\affiliation{Department of Engineering Science and Physics, The
College of Staten Island, City University of New York, Staten
Island, New York 10314}
\author{Nikolay Prokof'ev}
\author{Boris Svistunov}
\affiliation{Department of Physics, University of
             Massachusetts, Amherst, MA 01003}
\affiliation{Russian Research Center ``Kurchatov Institute'',
123182 Moscow }

\begin{abstract}
Spatially selective Ramsey spectroscopy is suggested as a method
for detecting the super-counter-fluidity  of 
two-component atomic mixture in optical lattice.
\end{abstract}


\maketitle

Recent advances in experimental studies of ultra-cold gases in
optical lattices \cite{Kasevich,Greiner} signal a major
breakthrough in the field of strongly-correlated quantum lattice
systems. Theoretical studies of ultracold atomic mixtures in
optical lattices have revealed a variety of non-trivial ordered
states \cite{Demler,KE,Paredes,KS,Demler2,Yip,SCF,1Ord}.
Developing experimental schemes able to resolve these states is of
crucial importance. Recently, Altman {\it et al}. \cite{Dem} has
considered generic possibilities of revealing pairing orders
through the density-density correlation properties. The proposed
in Ref.~\cite{Dem} scheme  for detecting pairing correlations
relies on analyzing noise in absorptive imaging. Yet, a direct
imaging of the non-single-component superflow as well as
of the non-trivial topological interplay between the
order parameters \cite{SCF,1Ord} (see also below)
are very desirable.
Here we observe that there exists a very simple (though not a
generic) method which immediately reveals the
super-counter-fluidity (SCF) \cite{KS,Demler2,SCF,1Ord} in the
systems formed out of the two {\it interconvertible} species (like
hyperfine states $|F=1,\,m_f=-1\rangle$ and $|F=2,\,m_f=1\rangle$
of $^{87}$Rb, which can be converted into each other by rf
radiation).

The SCF state occurs in a two-component lattice system at integer
total filling under the conditions of strong enough intra- and
inter-component repulsion. Basically, the state can be considered
as a condensate of pairs formed by particles of one component and
holes of another component. This picture is relevant to both
boson-boson, fermion-fermion and boson-fermion mixtures \cite{KS}.
In general, the components can be different elements. The
commensurability of the total filling factor guarantees that the
number of atoms of one component coincides with the number of
holes of another component, so that there is no non-paired atoms.
Under these conditions the net superfluid motion is impossible,
and only super-counter-flow can be realized \cite{KS}. The order
parameter associated with the SCF state is
\begin{equation}
\Phi_{SCF}=\langle \Psi^\dagger_B \Psi_A \rangle =|\Phi_{SCF}|\,
{\rm e}^{i\phi} \;
, \label{SCF1}
\end{equation}
where $\Psi_A$ and $\Psi_B$ are the field operators for the
component $A$ and $B$, respectively. The absence of condensate
of (bosonic) atoms
implies
\begin{equation}
\langle \Psi_A \rangle = \langle \Psi_B \rangle = 0
 \; . \label{SCF2}
\end{equation}

Another way of looking at the super-counter-fluid state---which is
especially relevant for the present consideration---is mapping it
onto the easy-plane ferro- (bosons) or anti-ferro- (fermions)
magnet \cite{KS,Demler2}. In general, the operators $S_z={1\over
2}\int (\Psi^{\dagger}_A\Psi_A- \Psi^{\dagger}_B\Psi_B)d{\bf x},\,
S_+=\int \Psi^{\dagger}_A\Psi_Bd{\bf x},\,\, S_-=S_+^\dagger$
represent the $su(2)$ algebra of the angular momentum operators
and thus can always be interpreted as  (pseudo-)spin operators
\cite{Kuk}. Eq.~(\ref{SCF1}) then means the easy-plane spin order
$\langle S_{\pm} \rangle \neq 0$.

In the spin terms,  the {\it equilibrium} ordering described by
the requirements (\ref{SCF1})-(\ref{SCF2}) is exactly equivalent
to the {\it non-equilibrium} ordering that arises in a {\it
normal} cloud of two-component mixture of, say, $^{87}$Rb atoms
created by the $\pi/2$ rf pulse \cite{Cornell} out of one
component. The spatially selective Ramsey spectroscopy (RS)
techniques has been successfully applied for detecting such
non-equilibrium spin order and its dynamics \cite{Cornell}. Hence,
one immediately concludes that the same technique should be
applicable for revealing the SCF phase. The only difference is
that the SCF order is formed spontaneously from two components,
which have no memory of each other. Thus, in contrast to the
situation \cite{Cornell}, only {\it one} $\pi/2$-pulse is needed.

Specifically, a short rf pulse produces the unitary transformation
\begin{equation}\label{U}
\displaystyle U={\rm e}^{-i(\omega'S_x + \omega''S_y)},
\end{equation}
where $\omega=\omega' + i\omega'' = \int \Omega (t)\, dt $ with
~$\Omega(t)$~ being the Rabi transition frequency, which enters
the Hamiltnonian $ H_P=(\hbar /2) [\Omega^*(t)S_+ + {\rm H.c.}]$
describing the effect of the rf-pulse---inter-conversion of the
components; $S_+=S_x + iS_y$.

Let us find the dominance of particles of the sort $A$ in the
final state: $\delta N= \langle N_A - N_b \rangle_{\rm fin}/2
\equiv  \langle S_z \rangle_{\rm fin}$. In terms of the initial
state, $\delta N=\langle S_z \rangle$~ after the pulse is given by
the relation $\delta N= \langle U^{\dagger}S_zU \rangle= \cos
|\omega|\langle S_z \rangle + \sin |\omega| (\omega' \langle
S_y\rangle -\omega'' \langle S_x\rangle )/|\omega|$ \cite{Kuk}.
For the $\pi /2$-pulse with real negative $\omega$ we, thus, have
\begin{equation}\label{omega}
\delta N= \int d{\bf x}\, |\Phi_{SCF}|\sin \phi \approx V \,
|\Phi_{SCF}| \, \sin \phi ~~~~~\mbox{($\pi /2$-pulse)} \; ,
\end{equation}
where $V$ is the volume of the system, and spatial uniformity of the
phase $\phi$ has been assumed. Given the fact that the
phase $\phi$ is arbitrary, we see that repeating the experiment
several times will result in the huge shot-to-shot noise ~$
|\delta N| \sim N$, provided the SCF is strong, that is, given by
the total number of particles ~$N$ as ~$ V|\Phi_{SCF}| \sim N$.

It is important to note that, similarly to the case
\cite{Cornell}, spatially resolved $\pi/2$-pulse will be able to
detect {\it local} phase profile, that is, super-counter-fluid
currents, including topological excitations \cite{KS,SCF,1Ord}. An
important ingredient is ability of controlling the circulation of
the SCF vortex. Eq.~(\ref{omega}) indicates that the local
dominance $\delta n=|\Phi_{SCF}|\sin \phi$ is sensitive to how the phase
winds. In general, should the winding of the SCF-phase exist,
initially uniform mixture will exhibit
a {\it phase-separation} pattern after the ~$\pi/2$-pulse. 
For example, in rotationally symmetric situation, the SCF phase
can be represented as ~$\phi= k \theta$, where ~$k=0, \pm 1, \pm 2,...$~ 
is the winding number \cite{note1}
and ~$\theta$~ is the polar angle in the plane, in which 
the SCF currents flow. Thus, the local dominance after the ~$\pi/2$-pulse
will be 
\begin{equation}\label{dn}
\delta n= |\Phi_{SCF}|\sin (k\theta).
\end{equation}
For ~$k\neq 0$, the domain 
boundaries are located along the lines given by $\theta = \pi m/k$,
with ~$m=0,1,.. (|k|-1)$.

 To exclude the possibility that the above-described interference
effect is actually due to the off-diagonal order in each {\it
separate} component, one needs just to make sure that there are no
single-component condensates, which can be easily done by the
absorption imaging technique \cite{Kett}.

In this report, we have concentrated on imaging the pure SCF
phase. However, properties of the mixture of two superfluids (2SF
phase discussed in Refs.~\cite{SCF,1Ord}), in which $\langle
\Psi_A \rangle\neq 0,\,\,  \langle \Psi_B \rangle \neq 0$ and,
yet, the SCF correlations remain strong (that is $ |\langle \Psi_A
\rangle| \approx |\langle \Psi_B \rangle| \ll |\Phi_{SCF}|$), are
quite unusual and deserve detailed experimental study. Such a
state can be realized by decreasing the lattice potential, so that
the SCF-2SF transition occurs \cite{1Ord}. One of the effects is
preserving circulation of the difference of the phases of the two
components, while the circulation of the sum is not preserved in
the SCF state \cite{SCF}. Accordingly, if, initially, the 2SF
state had one vortex of, e.g., sort A, in the SCF phase it will
become one SCF vortex, which can be viewed as a bound pair of {\it
1/2-vortices} of opposite circulations of the A and B components
(the circulation of the SCF vortex must be the same as of the
original vortex \cite{SCF}). Lowering the lattice potential, so
that the system returns back to the 2SF state, will result in
either reappearing of the vortex A or appearing of the anti-vortex
B. The outcome depends on which superfluid stiffness (A or B) is
smaller, so that the final vortex would have smaller energy. If
the energy of the vortex B is lower, this effect will be seen as
transferring circulation from A to B component by just cycling the
system through the 2SF-SCF transition.

Above, we have analyzed the case of bosonic mixture.
However, it is worth commenting on a
case of fermionic two-component mixture.
We note that the fermionic
SCF \cite{KS} corresponds to the
easy-plane anti-ferromagnetic order. Thus, imposing of
the $\pi/2$ pulse will not result in a global dominance
of one component. Instead, depending on the phase
of the SCF order parameter, the checkerboard order
will arise, with its contrast being modulated
by the original SCF phase. Resolving this effect
is, obviously, much more complicated issue and we will not
consider it here.  
 
Clearly, the above method cannot be applied for detecting the SCF
of {\it non-convertible} species such as different elements.
Furthermore, it will not work in the case of pairing of real atoms
(rather than atoms and holes).
Yet, it gives a unique opportunity to study new physical effects
in selected systems, which will, then, apply to the whole class
of the two-component quantum mixtures.


\begin{thebibliography}{99}
\bibitem{note1}
For a single SCF-vortex, $k=\pm 1$. In the case of the ring geometry,
multi-winding with $|k|>1$ is possible. If a group of several SCF-vortices
have separate cores, the above representation of the phase is valid
far from this group.
\end{thebibliography}
\end{document}